# Development of a novel neutron detection technique by using a boron layer coating a Charge Coupled Device

Juan Jerónimo Blostein[a,b,*], Juan Estrada[c], Aureliano Tartaglione[a,b], Miguel Sofo Haro[a,b], Guillermo Fernández Moroni[d,b] and Gustavo Cancelo[c]

[a] *Centro Atómico Bariloche, Instituto Balseiro-UNCuyo,*
  *Av. Bustillo 9500, R8402AGP, S.C. de Bariloche, Río Negro, Argentina*
[b] *CONICET, Argentina*
[c] *Fermi National Accelerador Laboratory*
  *Batavia, IL, 60510, USA*
[d] *Universidad Nacional del Sur*
  *Av. Colón 80, 8000FTN, Bahía Blanca, Pcia. de Buenos Aires, Argentina*

  *E-mail*: jeronimo@cab.cnea.gov.ar

ABSTRACT: This article describes the design features and the first test measurements obtained during the installation of a novel high resolution 2D neutron detection technique. The technique proposed in this work consists of a boron layer (enriched in $^{10}$B) placed on a scientific Charge Coupled Device (CCD). After the nuclear reaction $^{10}$B(n,$\alpha$)$^{7}$Li, the CCD detects the emitted charge particles thus obtaining information on the neutron absorption position. The above-mentioned ionizing particles, with energies in the range 0.5–5.5 MeV, produce a plasma effect in the CCD which is recorded as a circular spot. This characteristic circular shape, as well as the relationship observed between the spot diameter and the charge collected, is used for the event recognition, allowing the discrimination of undesirable gamma events. We present the first results recently obtained with this technique, which has the potential to perform neutron tomography investigations with a spatial resolution better than that previously achieved. Numerical simulations indicate that the spatial resolution of this technique will be about 15 μm, and the intrinsic detection efficiency for thermal neutrons will be about 3 %. We compare the proposed technique with other neutron detection techniques and analyze its advantages and disadvantages.

KEYWORDS: Neutron imaging; Neutron detection; Charge Coupled Device, position sensitive neutron detection.

## Contents

# 1. Introduction

Since the neutron discovery by Chadwick in 1932 [1, 2], the neutron detection techniques have been refined, and over the past two decades have seen remarkable progress [3-5]. This progress has been mainly motivated by the great capability of neutron techniques to study the structure, dynamics, composition and magnetization of condensed matter, in combination with the development and construction of high intense neutron sources. These sources are, nuclear research reactor and spallation sources, big facilities mainly dedicated to the neutron production [6]. Other applications that motivated the development of the neutron detection techniques are the instrumentation required for the operation of nuclear reactors [3], and the detection of concealed substances in port containers [7-12]. Due to the present international context, during the last decade this last application has become of importance for homeland security. Neutron detectors based on the use of helium-3, widely used for research during the last 30 years, were installed in the US borders in order to prevent the illegal traffic of radioactive materials [13]. This new application of the helium-3 based neutron detection, as well as the tens of thousands of liters that are being used in large science facilities in the US, have produced a deficit in the availability of helium-3 for research called ´helium-3 crisis´ [14]. For the reasons mentioned above, the development of alternative neutron detection techniques based on the use of different materials is of great interest nowadays for neutron scattering science continuity.

As thermal neutrons are not directly ionizing particles, some intermediate nuclear reaction is necessary in its detection process. One example widely used are the helium-3 proportional counters, in which the nuclear reaction is produced in the helium-3 used to filling the detector active volume. Other example of gas detectors are the proportional counters based on $BF_3$, in which the nuclear reaction is produced in the $^{10}B$ isotope. The mentioned alternative has two disadvantages: $BF_3$ is very corrosive and extremely toxic. For these reasons, the $BF_3$ detectors life-time is limited and its handling and storage must be strictly controlled [15].

In this context, during the last years different detection techniques were developed for position sensitive neutron detection. These techniques include those based on the use of different scintillator materials coupled to Wavelength-Shifting Fibres (WSFs) and Photo-Multiplier Tubes (PMTs) [16, 17], $^{10}B$-lined detectors coupled to Multi-Wire Proportional Chambers [18], straws tubes built with an inner coating of $^{10}B_4C$ and Multi-Grid Detectors [19, 20, 18], as well as those techniques that use Micro Chanel Plates (MCP) coupled to charge-coupled devices (CCD) [21-24]. The advantages and disadvantages of the mentioned techniques are analyzed in reference [18]. In the case of the technique based on MCP devices, the neutron absorption is produced in the boron contained in the MCP structure. The charged particles emitted after the nuclear reaction produce electrons that are multiplied in the MCP channels. Then, the electron avalanche is detected by the CCD. With this technique the obtained spatial resolution is approximately 60 micrometers, which is essentially given by the channel spacing [21-24].

The CCD was invented in 1969 at AT&T Bell Labs by Willard Boyle and George E. Smith [25-26]. The essence of the design is the capability to transfer charge onto a semiconductor surface from one storage capacitor to the next. Scientific CCDs, originally developed for photon detection, have been extensively used in ground and space-based astronomy, X-ray imaging and other particle detection applications [27]. The combination of high detection efficiency, low noise, good spatial resolution, low dark current and high charge transfer efficiency results in an excellent performance for detection of ionizing particles [28].

The purpose of this work is to introduce a novel neutron detection technique that employs the formerly mentioned CCD characteristics in combination with a proper material that, after a neutron induced nuclear reaction, emits charged particles that can be easily recorded by the CCD. The proposed technique is based on a scientific CCD cover with a boron layer.

The advances performed in its implementation and the obtained experimental results are shown together with numerical results that indicate that the proposed technique will be very useful, especially for those applications where high spatial resolution in the neutron detection position is required. Among these applications, it can be mentioned the neutron imaging formation in transmission experiments, technique of interest for visualisation of distributions of chemical elements in different materials [29], crystallographic phases and crystalline orientations [30], and several application in Material Science [31], Biology, Geology, etc. [29, 32-34].

## 2. The CCD used

Figure 1 shows the CCD used in this work. This CCD was developed for the DECam wide field imager that is currently under construction [35]. The CCD was built by the Lawrence Berkeley National Laboratory (LBNL) [36], and extensively characterized at Fermilab for the DECam project [37]. The CCD used in this work is 250 μm thick, fully depleted, back-illuminated device fabricated on high-resistivity silicon. It has 2k x 4k pixels of 15 μm x 15 μm each, resulting on an effective area of approximately 3 x 6 cm$^2$. The CCD was placed inside a vacuum box, and cooled by using liquid nitrogen. The measurements were made with the CCD working at approximately 135 K to avoid any spurious generation of charge. At this temperature the dark current contribution is less than a 1e-/pixel/hour and has a negligible effect on the alpha detection [38].

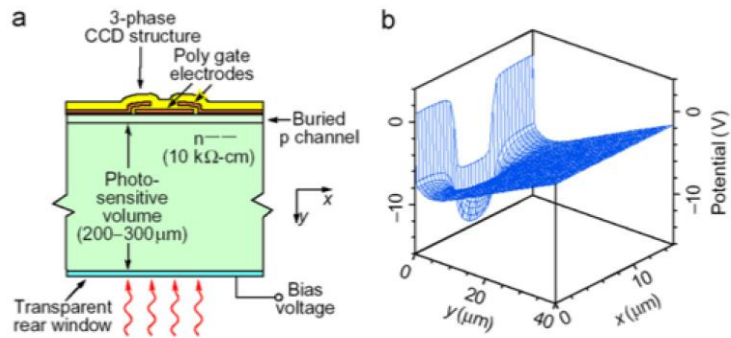

*Figure 1: (a) Pixel cross section of a 250 μm thick CCD developed at Lawrence Berkeley National Laboratory. (b) The electrostatic potential (V) generated through three gated phases is shown as function of depth (y axis) and one of the lateral directions (x axis). The generated charge is stored in the potential well.*

Figure 2 shows an image obtained by exposing the CCD active surface to different ionizing radiations. The recorded event shape can be used to identify the incident particle.

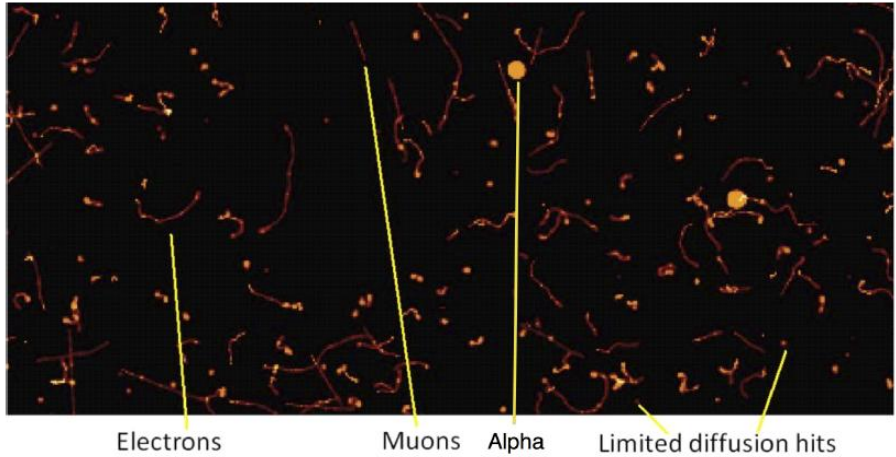

*Figure 2: Events recorded with the CCD exposed to different ionizing radiations. The limited diffusion hits are the result of the interaction between the silicon and X-rays from a $^{55}$Fe source.*

The typical reading time of the charge cumulated in each pixel is about 3 µs, thus producing a total reading time of about 24 seconds for each image. During the reading time, the device is still exposed to ionizing events, and therefore this period is considered as part of the total exposure time. It must be noted that the events recorded during the reading process can modify the obtained image.

## 3. The proposed technique

The technique proposed in this work consists of a boron layer enriched in $^{10}$B coating a scientific CCD. As Figure 3 shows, the object to be analysed is placed between the neutron source and the CCD, so some neutrons are absorbed or scattered by the object, while others can reach the boron layer. As depicted in Figure 4, after the nuclear reaction $^{10}$B(n,α)$^{7}$Li takes place, some of the emitted charged particles can leave the boron layer and be detected by the CCD. In the image recorded after the CCD readout it is possible to observe the transmitted neutron beam shape. The neutron detection active area is defined by of the CCD active dimensions.

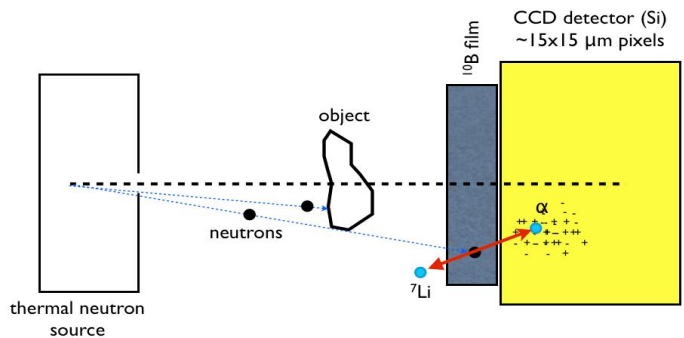

*Figure 3: Experimental set up. The sample is placed in a collimated neutron beam.*

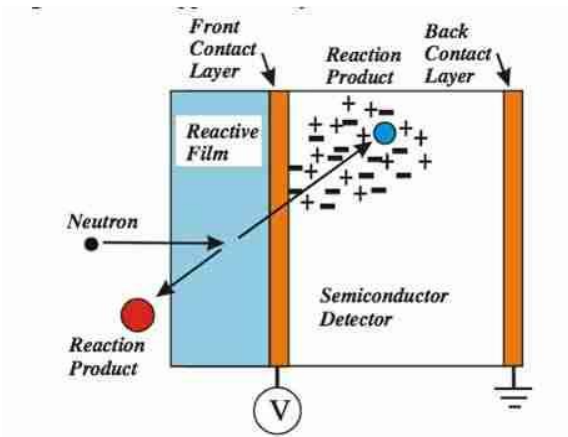

*Figure 4: A boron layer is placed on the surface of a CCD. The charged particles ($\alpha$ and $^{7}Li$) emitted in the boron nuclear reaction are detected in the CCD.*

## 4. Preliminary test with alpha particles

In the proposed neutron imaging technique the alpha particles produced in the above mentioned nuclear reaction are directly detected by the CCD. As a preliminary test we expose the CCD to an $^{241}$Am α source (approximately 5.5 MeV). Figure 5 shows one of the obtained images.

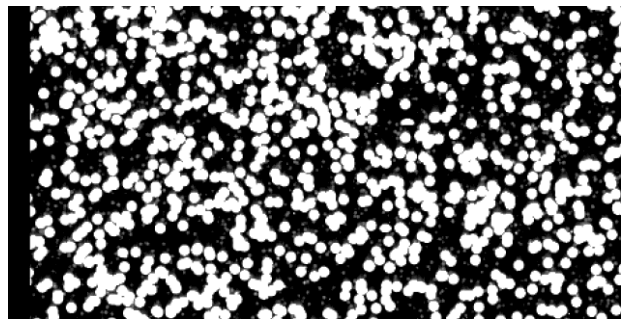

*Figure 5: Image obtained with the CCD exposed to a $^{241}$Am source.*

The alpha events are clearly recorded as spots of about 10 pixels diameter. The gamma events (with energies lower that 60 keV) are recorded as smaller spots and small tracks. Figure 6 shows the corresponding energy spectrum obtained by adding all the pixel charge of each event. It shows two different ranges in the energy of the detected particles, making a clear distinction between 5.5 MeV alpha particles and photons with energies lower than 60 keV.

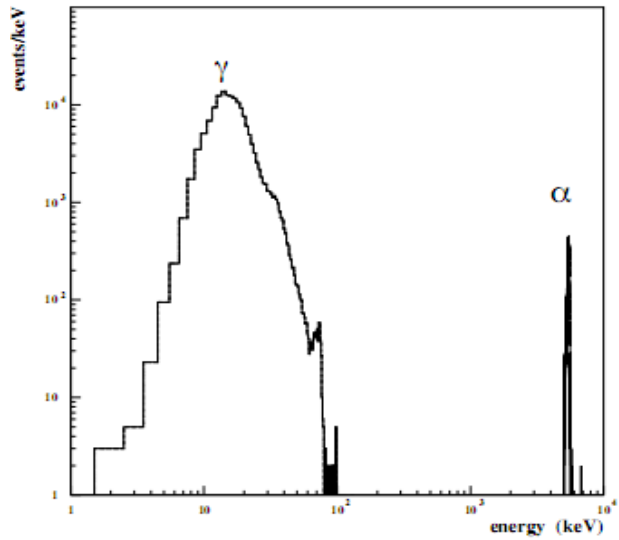

*Figure 6: Energy measured by the CCD when exposed to a $^{241}$Am source.*

The detected alpha particle energy depends on the cluster size. Figure 7 shows the experimental measurements of the relationship between these magnitudes. The behavior observed in Figure 7 is mainly due to a plasma effect produced by the charged particles in the silicon bulk described in reference [39]. In the image processing, this relationship can be used to discriminate some events not produced by charged particles emitted by the boron layer. Another characteristic that can be used to validate one event during the image processing is the event symmetry. Muons and gamma rays normally produce non symmetric events; the second order momentum of the charge in the horizontal axis is different from that of the vertical axis.

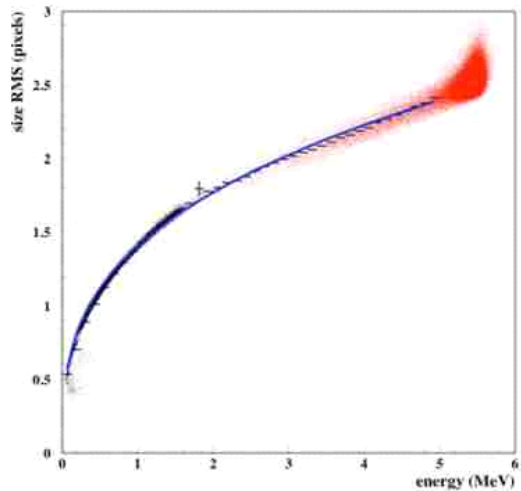

*Figure 7: Energy dependence on the cluster size for alpha particles. Red points were obtained with the $^{241}$Am source, and black points with $\alpha$ particles coming from (n,$\alpha$) reactions.*

## 5. Experimental set up

For this preliminary test, boron was not directly placed on the silicon CCD surface, but on a 2 mm thick aluminium plate (5 x 5 cm$^2$ size), shown in Figure 8. The $^{10}$B layer was deposited by electro-deposition, and its thickness (2.07±0.08 μm) was measured by neutron transmission experiments in the thermal neutron energy range by using the 25 MeV Bariloche electron LINAC accelerator and the time-of-flight method [40-41]. The boron surface was placed at approximately 1.1 mm from the CCD.

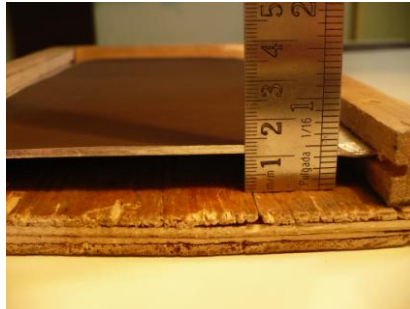

*Figure 8: Aluminium plate with a $^{10}B$ layer on its surface.*
*The borated face was placed near the CCD surface (less than 1 mm).*

In Figure 9 we show details of the CCD and its connections, as well the box where the CCD is placed and cooled. The CCD was fully depleted by using a 40V substrate voltage.

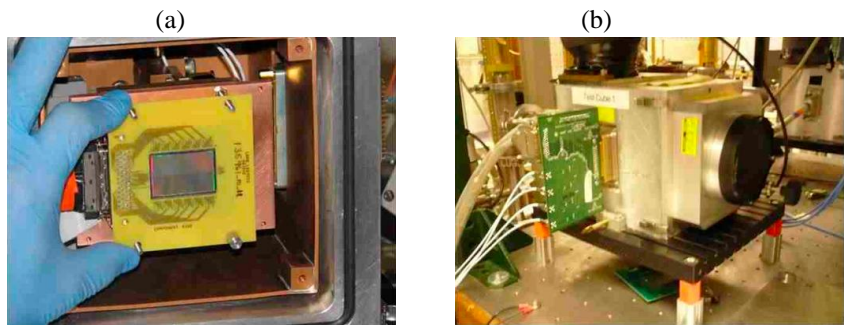

*Figure 9: (a) CCD placed in its dewar without the borated surface and without the dewar cover.*
*(b) CCD dewar with its cover.*

For this preliminary test we use as a sample a 1 mm thick cadmium plate with a cross shape hole placed in contact with the aluminium plate inside the dewar. A picture of this object is shown in Figure 10. It is 6.8 cm long and 4.1 cm wide.

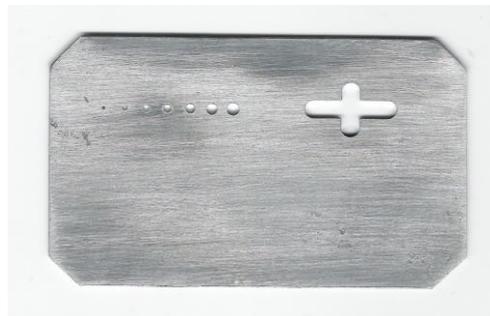

*Figure 10: Cd plate with a cross and different holes, employed as a sample.*

# 6. Results

## 6.1 Experimental results

The neutron image presented in Figure 11 was obtained by using a 100 mCi $^{252}$Cf neutron source placed at 10 cm from the CCD (outside the dewar), a 3.8 cm thick polyethylene slab as neutron moderator, and the experimental set-up described in the previous section. The exposure time was 400 minutes, and the reading time 24 seconds. Figure 11 is presented in an inverted gray scale; the white background corresponds to regions without ionizing process. The black dots are due to alpha and lithium particles produced by nuclear reactions in the boron by those neutrons that are not absorbed in the cadmium plate. In Figure 11 it is possible to see several black dots inside the cross shape hole, as well as in other regions where the cadmium did not cover the boron (upper and lower sides of the image). Besides, some black dots are also observed in the area covered by the cadmium sheet. This is so because some epithermal neutrons can pass through the cadmium and be absorbed by the boron. In the area covered by the cadmium the observed dots density is significantly lower than that of the uncovered area.

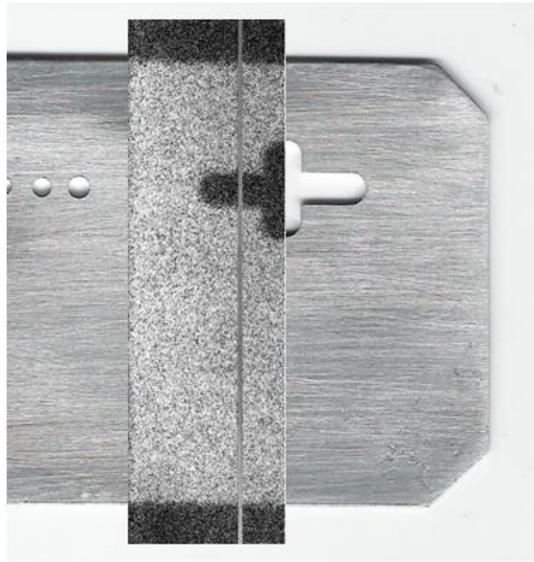

*Figure 11: Neutron image obtained with this technique (central part). The background of this figure is a photo of the cadmium plate (zoom of Figure 9). For the neutron image an inverted gray scale was used. The regions where the CCD detects more alpha and lithium events are shown in black.*

## 6.2 Resolution and efficiency estimation by Monte Carlo simulations

It is worth noticing that for this preliminary experiment the boron layer was deposited on an aluminum surface placed at about 1 mm from the silicon surface. The intrinsic detection efficiency of an ideal detection system, with the boron layer directly coating the CCD silicon surface, was estimated by means of Monte Carlo simulation by using the MCNP code. In order to validate the employed code, the energy spectrum of the alpha particles emitted by the boron layer in presence of thermal neutron is compared with that obtained from the simulation. The measurements were performed by using a silicon detector connected to a charge sensitive preamplifier. In Figure 12 the experimental pulse height spectrum is compared with that obtained from the Monte Carlo simulation.

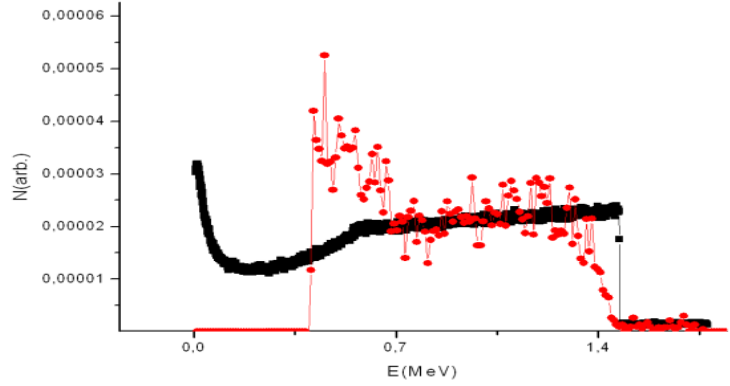

*Figure 12. Red: Experimental energy spectrum of the charged particles emitted by boron layer after the absorption of thermal neutrons. Black: Simulated energy spectrum taking only into account alpha particles.*

The employed MCNP version does not take into account the $^7$Li contribution. This is the reason for the difference observed in Figure 12 below 0.7 MeV. The simulation does not take into account a gold layer deposited on the silicon detector used for this measurements. This is the reason for the difference observed at about 1.5 MeV.

In a real experiment by employing the technique introduced in this work, a lower level discrimination (LLD) must be selected according to the observed background level. The background level has three components: the CCD dark current noise, the electronic read-out noise, as well as the radiation file present at de detection position (commonly due to gamma rays). Only events with energy greater than LLD should be counted as neutrons. The detection efficiency is calculated as the area under the energy spectrum. For this reason, the optimum boron thickness depends on the LLD. For a very thin boron layer most of the neutrons are transmitted without interaction, and for a very thick one the α and $^7$Li do not reach the CCD. Figure 13 shows the optimum boron thickness, $T_{opt}$, as a function of LLD obtained from the simulations without taking into account the $^7$Li contribution. Figure 14 shows detection efficiency for $T_{opt}$ as a function of LLD.

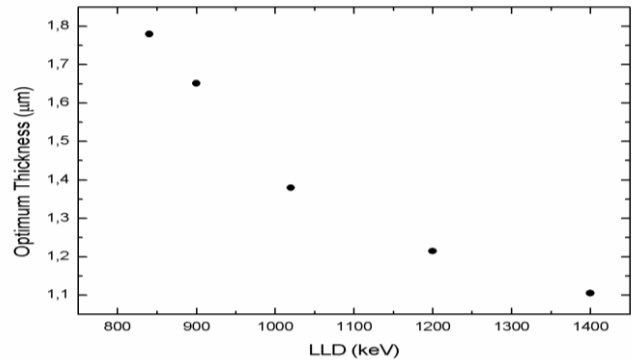

*Figure 13. Optimum boron thickness as a function of the lower level discrimination (LLD).*

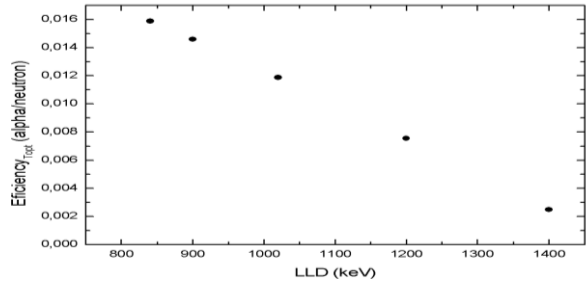

*Figure 14. Neutron detection efficiency for the optimum boron thickness presented in Figure12.*

Finally, in order to estimate the spatial resolution of the proposed technique, we use the SRIM2011 code to simulate the tracks of α particles in $^{10}$B and Si. We assume an incident α beam of 1780 keV in the horizontal direction. The results are shown in Figure 15.

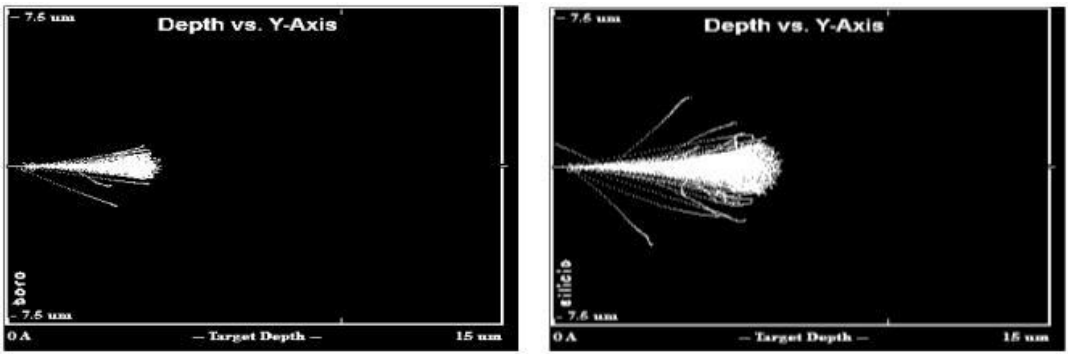

*Figure 15. Simulation of α tracks in $^{10}$B (left) and Si (right) by using the SRIM2011 code assuming an incident α beam of 1780 keV in the horizontal direction.*

We observed a mean α range of about 4.4 μm in $^{10}$B, and 6.4 μm in Si. It is worth noticing that the charged particles are emitted isotropically after the nuclear reaction, and that the CCD cannot detect the ion incident angle. Therefore, the ion ranges analysed above are useful to estimate this technique spatial resolution. From the maximum range formerly mentioned, it is possible to estimate that this technique spatial resolution will be approximately 15 μm, similar to the CCD pixel spacing employed in this work.

## 7. Conclusion

The preliminary tests of the technique introduced in this work have been successful, and the first neutron images have been obtained. The unwanted events (gammas, X rays, muons, fast neutrons, etc.) can be mostly discriminated by analysing the energy events, as well as their distinctive trace on the CCD. Another characteristic that can be used to validate each recorded event as one neutron is the relationship existing between the event diameter and the charge collected shown in Figure 7. From simulations that do not take into account the $^{7}$Li contribution, assuming a thermal neutron beam with its axis normal to the CCD surface, the maximum neutron detection efficiency is about 1.6%. Including the $^{7}$Li contribution we estimate a total efficiency for thermal neutron of about 3% for a low background condition. The detection efficiency could be increased by placing the CCD surface in a different angle. In the mentioned configuration the neutron absorption probability increases without affecting the probability that the charged particles escape from the boron layer. This device can be used for different neutron imaging applications, especially in those cases were high spatial resolution is required. In order to improve the spatial resolution the following step will be to place the boron layer directly on the CCD surface. From Monte Carlo simulations we have estimated a spatial resolution for this technique of about 15 μm, which is better than the best spatial resolution of MCPs devices (approximately 65 μm) [21-24]. For this reason, the proposed technique will be especially useful for those applications where high spatial resolution in the neutron detection position is required. One example is the neutron imaging formation in transmission experiments. Other examples are the beam alignment and the sample positioning in different kind of experiments, which could be performed by placing the detection system in the neutron beam. This technique has the advantage of not requiring an MCP device. In addition, the spatial resolution is practically independent of the angle between the incident neutron and the borated surface. Moreover, large active areas could be obtained at low cost. Preliminary experimental tests indicate that this technique could be also employed by using boron coating a CMOS device. CCD and CMOS devices with timing resolutions of about 1 μs are used in neutron imaging applications [21-24] and are commercially available. The mentioned devices will also allow to use this technique in neutron time-of-flight measurements. Detection efficiencies for thermal neutrons greater than 3% could be reached by employing a multi layer arrange of boron-CCD or boron-CMOS devices.

## Acknowledgments

We acknowledge Gregory Derylo, Kevin Kuk and Heman Caese for technical support at Fermilab (USA), and Marcelo Miller, Luis Capararo and Yamil Moreira for the borate aluminium plate preparation at CNEA (Argentina). This work was partially supported by ANPCyT (Argentina) under project PICT 2011-0534 and by CONICET (Argentina) under the project PIP 2011-0552.

## References

[1] J. Chadwick, *Possible Existence of a Neutron*, Nature 129, 312 (1932).

[2] J. Chadwick, *The Existence of a Neutron*, Proceedings of the Royal Society of London A136, 692-708 (1932).

[3] G.F. Knoll, *Radiation Detection and Measurement*, J. Wiley & Sons (1979-89-2000).

[4] *New Developments in Radiation Detectors*, Neutron News, 16(4) (2005) 13-21.

[5] A.J. Peurrung, *Recent developments in neutron detection*, Nucl. Instrum. and Meth. A, Volume 443, Issues 2–3, 1 (2000), 400–415.

[6] The ISIS Facility, Annual Report 2002–2013, ISSN 1358–6254.

[7] T. Gozani, *The Role of Neutron Based Inspection Techniques in the Post 9/11/01 Era*, Nucl. Instr. Meth. B 213 (2004) 460-463.

[8] L.J. Jones *et al.*, *Detection of Shielded Nuclear Material in a Cargo Container*, Nucl. Instr. Meth. A 562(2), 01/2006, 1085-1088.

[9] P. Kerr *et al.*, *Active Detection of Small Quantities of Shielded Highly-enriched Uranium Using Low-dose 60-keV Neutron Interrogation*, Nucl. Instr. Meth. B 261 (2007) 347-350.

[10] J.M. Hall *et al.*, *The Nuclear Car Wash: Neutron Interrogation of Cargo Containers to Detect Hidden SNM*, Nucl. Instr. Meth. B 261 (2007) 337–340.

[11] K.A. Jordan and T. Gozani, *Pulsed Neutron Differential Die Away Analysis for Detection of Nuclear Materials*, Nucl. Instr. Meth. B 261 (2007) 365-368.

[12] A. Tartaglione, F. Di Lorenzo, R.E. Mayer, *Detection of Thermal-Induced Prompt Fission Neutrons of Highly-Enriched Uranium: A Position Sensitive Technique,* Nucl. Instr. and Meth. B 267 (2009) 2453-2456.

[13] D. A. Shea and D. Morgan. *The Helium-3 shortage: Supply, demand, and options for congress*. Technical Report R41419, Congressional Research Service, 2010.

[14] A. Cho. *Helium-3 shortage could put freeze on low-temperature research.* Science, 326(5954), 2009.

[15] T. Wilpert, *Boron Triouride Detectors*, Neutron News, 23(4), 2012.

[16] N. G. Rhodes, *Scintillation Detectors*, Neutron News, 23(4), 2012.

[17] T. Nakamura *et al*. *A large-area two-dimensional scintillator detector with a wavelength-shifting fibre readout for a time-of-flight single-crystal neutron difractometer*, Nucl. Instrum. and Methods A, 686(0), 2012.

[18] *$^{10}B_4C$ Multi-Grid as an alternative to $^3He$ for Large Area Neutron Detectors*, Correa Magdalena, Jonathan ; Guerard, B. (dir.) ; Campo Ruiz, J.J. (dir.), Universidad de Zaragoza, (http://zaguan.unizar.es/record/9939/files/TESIS-2013-009.pdf)

[19] J.L. Lacy et al., *Novel neutron detector for high rate imaging applications*. In Nuclear Science Symposium Conference Record, 2002. IEEE, volume 1, 2002.

[20] A. Athanasiades et al. *Straw detector for high rate, high resolution neutron imaging*. In Nuclear Science Symposium Conference Record, 2005. IEEE, volume 2, 2005.

[21] A. S. Tremsin, J. V. Vallerga, J. B. McPhate, O. H. W. Siegmund, W. B. Feller, L. Crow, R. G. Cooper, *On the possibility to image thermal and cold neutron with sub-15 μm spatial resolution*", Nucl. Instr. and Meth. A 592 (2008) 374.

[22] A. S. Tremsin, J. B. McPhate, J. V. Vallerga, O. H. W. Siegmund, J. S. Hull, W. B. Feller, E. Lehmann, *High-resolution neutron radiography with microchannelplates: Proof-of-principle experiments at PSI*, Nuclear Instruments and Methods in Physics Research A, 605 (2009) 103–106.

[23] Anton S. Tremsin, , W. Bruce Feller, R. Gregory Downing, *Efficiency optimization of microchannel plate (MCP) neutron imaging detectors. I. Square channels with 10B doping,* Nuclear Instruments and Methods in Physics Research A 539 (2005) 278–311

[24] A. S. Tremsin, J. B. McPhate, J. V. Vallerga, O. H. W. Siegmund, J. S. Hull, W. B. Feller, E. Lehmann, *Detection efficiency, spatial and timing resolution of thermal and cold neutron counting MCP detectors*, Nucl. Instr. and Meth. A 604 (2009) 140.

[25] G.E. Smith, THE INVENTION AND EARLY HISTORY OF THE CCD, Nobel Lecture, December 8, 2009.

[26] W.S. Boyle, CCD – AN EXTENSION OF MAN'S VISION, Nobel Lecture, December 8, 2009.

[27] J. Estrada, et al., *Prospects for a Direct Dark Matter Search Using High Resistivity CCD Detectors*, arXiv:0802.2872v3, 2008.

[28] J.R. Janesick, *Scientific Charge-Coupled Devices*, SPIE Press, Bellingham,WA, 2001.

[29] A.S. Tremsin et al., IEEE Transactions on Nuclear Science, **52**, 1739 (2005).

[30] W. Kockelmann, G. Frei, E.H. Lehmann, P Vontobel, J.R. Santisteban, Nucl. Instr. Meth. **A578**, 421 (2007).

[31] N Kardjilov, E Lehmann, E Steichele, P Vontobel, Nucl. Instr. Meth A527, 519–530.

[32] EH. Lehmann et al, Nucl. Instr. Meth. A603, 429 (2009).

[33] E. H. Lehmann et al., Nuclear Instruments and Methods in Physics Research A 603 (2009) 429-438.

[34] W. Kockelmann et al., Nuclear Instruments and Methods in Physics Research A 578 (2007) 241-234.

[35] Dark Energy Survey Collaboration, The Dark Energy Survey, astro-ph/0510346, 2005.

[36] S.E. Holland, D.E. Groom, N.P. Palaio, R.J. Stover, M. Wei, IEEE Transactions on Electron Devices ED-50 (2003) 225.

[37] C. Abbott, et al., *Ground-based andairborne instrumentation for astronomy*, in: I.S. Mc Lean, M. Iye (Eds.), Proceedings of the SPIE, vol. 62 69, 2006.

[38] H.T. Diehl, et al., *Characterization of DECam focal plane detectors,* In: Dorn, D.A., Holland, D. (eds.) High Energy, Optical and Infrared Detectors for Astronomy III, Proceedings of the SPIE, vol. 7021, p. 70217 (2008).

[39] J. Estrada, J. Molina, J.J. Blostein, and G. Fernández, *Plasma effect in silicon charge coupled devices (CCDs)*, Nuclear Instruments and Methods in Physics Research A 665 (2011) 90–93.

[40] *Search for Anomalous Effects in $H_2O/D_2O$ Mixtures by Neutron Total Cross Section Measurements*, J.J. Blostein, J. Dawidowski, S.A. Ibáñez and J. R. Granada, Physical Review Letters, 90, 105302 (2003).

[41] *Total cross sections of benzene at 90 K and light water ice at 115 K*, L. Torres, J.R. Granada, J.J. Blostein, Nucl. Instruments and Methods in Physics Research B, 251 304-305, 2006.